\documentclass[a4paper,fleqn,usenatbib]{mnras}

\usepackage{newtxtext,newtxmath}

\usepackage[T1]{fontenc}
\usepackage{ae,aecompl}

\usepackage{graphicx}	
\usepackage{amsmath}	
\usepackage{amssymb}	
\usepackage{gensymb}
\usepackage{multirow}

\title[Analysis of the June 2, 2016 bolide event over Arizona]{Analysis of the June 2, 2016 bolide event over Arizona}

\author[Cs. Palotai et al.]{Csaba Palotai,$^{1}$\thanks{E-mail: cpalotai@fit.edu (CsP)}
Ramanakumar Sankar,$^{1}$
Dwayne L. Free, $^{2}$ 
J.\ Andreas Howell,$^{3}$
\newauthor Elena Botella$^{1}$
and Daniel Batcheldor$^{1}$
\\
$^{1}$Department of Physics \& Space Sciences, Florida Institute of Technology, 150 W. University Blvd., Melbourne, FL 32901, USA\\
$^{2}$ Spalding Allsky Camera Network, SkySentinel, LLC, 958 Shaw Circle, Melbourne, FL 32940, USA\\
$^{3}$ Department of Online Science, Florida Institute of Technology, 150 W. University Blvd., Melbourne, FL 32901, USA\\
}

\date{Accepted XXX. Received YYY; in original form ZZZ}

\pubyear{2017}

\begin{document}
\label{firstpage}
\pagerange{\pageref{firstpage}--\pageref{lastpage}}
\maketitle

\begin{abstract}
On June 2, 2016 at 10h56m UTC, a  $-20.4 \pm 0.2$ magnitude superbolide was observed over Arizona. Fragments were located a few days later and the meteorites were given the name Dishchii'bikoh. We present analysis of this event based on 3 cameras and a multi-spectral sensor observations by the SkySentinel continuous fireball-monitoring camera network, supplemented by a dash cam footage and a fragmentation model.  The bolide began its luminous flight at an altitude of $100.2 \pm 0.4$ km at coordinates $\phi = 34.555 \pm 0.002\degree$N planetographic latitude and $\lambda = 110.459 \pm 0.002\degree$W longitude, and it had a pre-atmospheric velocity of $17.4 \pm 0.3$ km/s.  The calculated orbital parameters indicate that the meteoroid did not belong to any presently known asteroid family.  From our calculations,  the impacting object had an initial mass of $14.8 \pm 1.7$ metric tonnes with an estimated initial diameter of $2.03 \pm 0.12$ m. 
\end{abstract}



\section{Introduction}
At 10h 56m 27s UTC on June 2, 2016 a bright fireball was observed over Arizona.  The American Meteor Society received 421 reports about this fireball, most of them from Arizona but people also witnessed this event in Utah, New Mexico, California, Texas, Colorado and Nevada.  Sonic booms from the bolide were heard across the greater Phoenix area.  Videos of the bolide from dash cams and security cameras appeared on YouTube and various media outlets.  After sunrise, videos captured the dust trail that the impactor left behind.   

Camera and satellite observations of meteors have been used for decades in order to help to determine the mass, trajectory and orbital parameters of the impacting body.  Various satellites, NASA's All Sky Fireball Network and the Lowell Observatory Cameras for All-Sky Meteor Surveillance (LO-CAMS) also recorded the Arizona bolide event.  The Arizona Geological Survey's seismic network picked up a signal near Payson that was consistent with an airburst event and according to the agency it marked the explosion of the asteroid. This was the largest observed bolide event over the continental United States since March 2010 that was reported by the Center for Near Earth Object Studies (CNEOS)\footnote{{\texttt https://cneos.jpl.nasa.gov/fireballs/}, accessed December 2, 2017}, and the fifth largest since 1988 when the agency began recording fireball data from US Government sensors.

\begin{figure}
	\includegraphics[width=\columnwidth]{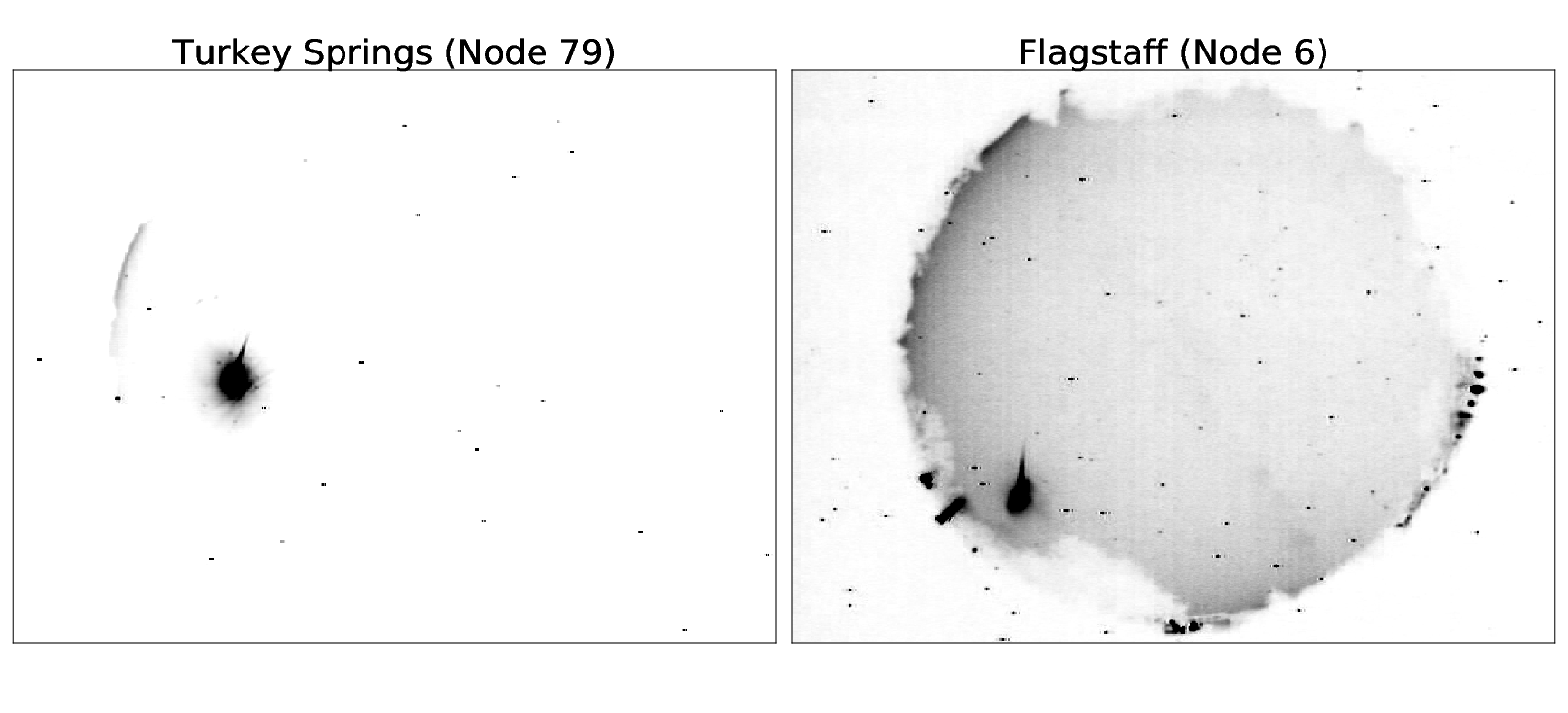}
	\caption{Composite images of the bolide from Node 6 (Flagstaff, AZ) and Node 79 (Turkey Springs, AZ). Only the first 3s were used from each video to create the composites due to  saturation of subsequent frames. The event itself continues for about an additional $4$s and brightens by several orders of magnitude. } 
	\label{fig:composite}
\end{figure}

The bolide was also observed by multiple nodes of the SkySentinel's Spalding Allsky Camera Network (Figure~\ref{fig:composite}) and the recorded data allow us to characterize some of the physical properties and the likely origin of the impacting body.  Here, we present our analysis of the available SkySentinel observations, coupled with complementary data and modeling tools.  We also compare our inferred results with data reported by other sources.      

\section{Methodology and Instrumentation}
\subsection{The SkySentinel Network}
Mr.\ R.E. Spalding (1936-2017, Sandia National Labs) developed the Allsky Camera System to monitor, track, and analyze large meteor events in order to provide ``ground-truth'' to assist both science (NASA) and treaty monitoring (Nuclear-Test-Ban Treaty Organization - CTBTO) operations in confirming the impact of large meteor fireballs (bolides) in Earth's atmosphere.  The collected data is also used to support the refinement of the energy calculations of those events, as well as, to improve trajectory calculations and orbit determination for the impacting bodies.  Another important goal of the project was to develop a companion instrument, MultiSpectral Radiometer (MSR) for comparison to the Allsky cameras, and government-collected data, in order to improve the diagnostic capability of SkySentinel.

The network consists of a large number of wide-angle view cameras at various sites throughout the continental United States and other countries and it also includes the infrastructure that permits the archival of the observational data that is available for processing and analysis for the scientific community. Software tools were developed and added for calibration, removal of detector effects and anomalies, automatic event detection and correlation among stations, and automatic trajectory computation. 

Mr. R. E. Spalding formally established the current SkySentinel Allsky Camera Network that was transitioned to SkySentinel, LLC and is now operated as a Joint Florida Institute of Technology and SkySentinel, LLC Science Education and Research Program.  This paper is a demonstration of improvements made to SkySentinel under the Joint Program.  The current system was renamed the Spalding Allsky Camera Network in honor of its founder.  Data for recorded events and details on the camera system can be found at \texttt{www.goskysentinel.com}.


\subsection{Camera System Parameters}

The 2 June 2016 Arizona bolide event was detected by seven SkySentinel nodes. Table~\ref{tab:nodes} lists the SkySentinel nodes that observed the event. These nodes include a Sony HB-710E Starlight B/W CCD camera coupled with a Fujinon YV22X14ASA 3.4mm lens.
 The cameras are sensitive down to about $+2$ visual magnitude using 8.5 second image composites. For individual frames recorded at 30 frames per second, the cameras are sensitive to just above magnitude $0$.

The node cameras are enabled throughout the night and the video feed goes directly into the WSentinel tracking software which automatically triggers on motion/threshold crossing and saves 1s of presamples and the entirety of the event in four files: composite .jpg image, .mp4 video (640x480 @ 30fps; 8 bits per pixel), .csv spreadsheet of time stamped azimuth/elevation, sum of pixels triggering, sum of energy above threshold per pixel, X/Y pixel map of the event and a .txt file of the start/stop time of the event recording. WSentinel also takes an hourly composite .png image  (adjustable, 8.5s minimum) that is used to calibrate the camera by star mapping tie points.


\subsection{Astrometric Calibration of SkySentinel Sensors}
SkySentinel nodes report positional data that are expressed in Cartesian coordinates $(x, y)$.
During the calibration of the data we convert each SkySentinel camera's reported $(x, y)$ measurements into horizon coordinates $(z, a)$, where $z$ is the zenith distance and $a$ is the azimuth. The image calibration is accomplished by fitting a 10-parameter mathematical model to observations of calibration stars in each node's CCD imagery. Table~\ref{tab:nodes} lists the SkySentinel nodes that were calibrated, the dates of the SkySentinel images used for calibration, and the number of calibration points (stars) employed.

\begin{table*}
	\begin{center}
		\caption{Planetographic coordinates for the SkySentinel nodes that observed the event, and the dates and number of stars used for the astrometric calibration of the pixel positions from the cameras. Nodes 5, 6, 8, 37 and 79 are used in this study.}
		\label{tab:nodes}
		\begin{tabular}{ccccccc}
			\hline
			Node & Location & Imagery Dates	& No. Cal Stars & Latitude ($\degree$) & Longitude ($\degree$) & Altitude (m)\\
			\hline
			\hline
			1 & Las Cruces, NM 	& Jun 1-4	& 130  & 32.281 & -106.754 & 1191 \\
			2 & Las Cruces, NM 	& May 31-Jun 2 & 54 & 32.281 & -106.754 & 1191              \\
			5 & Alberquerque, NM & May 29 - Jun 18 & 102 & 35.152 & -106.556 & 1600 \\
			6 & Flagstaff, AZ 	& May 31-Jun 2 & 124 & 35.200 & -111.655 &  2106            \\
			14 & Los Alamos, NM 	& May 31-Jun 4 & 56 & 35.887 & -106.277 & 2177    \\
			37 & Parker,	AZ 		& May 31-Jun 2 & 109  & 34.144 & -114.291 & 128        \\
			79 & Payson,	AZ 		& May 31-Jun 3 & 120  & 34.233 & -111.301 & 1514         \\
			8$^*$ & Alberquerque, NM & - & - & 34.951 & -106.460 & 1958 \\
			\hline
		\end{tabular} \\
	\end{center}
	\begin{flushleft}
		$^*$ MultiSpectral Radiometer (MSR)
	\end{flushleft}
\end{table*}

The method used to perform the calibration was based on a geometric solution originally developed by \citet{borovicka1995} to calibrate emulsion-based plates from all-sky cameras.  Borovi{\v c}ka's complete solution, which employed 13 calibration parameters, achieves a residual error of just 0.015 degree. The solution's main drawback is that it requires a ``large number'' of calibration stars, including stars at large zenith angles, which are ``seldom available''.

The Borovi{\v c}ka solution uses a coordinate system that differs in an important way from SkySentinel's coordinate system. To distinguish between them, SkySentinel coordinates will henceforth be denoted using lower-case letters $(x,y)$, and upper-case letters will denote Borovi{\v c}ka coordinates $(X,Y)$. The difference is that coordinates are interchanged, i.e., $(X,Y)$ correlates to $(y,x)$. For example, $(x=320, y=240)$ in SkySentinel coordinates is the same point as $(X=240, Y=320)$ in Borovi{\v c}ka coordinates. 

On the other hand, the two coordinate systems are alike in that they both define the center of the upper left pixel to be the origin of the coordinate system $(0,0)$.  For SkySentinel cameras aligned approximately with North up, the origin $(0,0)$ is the center of the pixel in the Northeast corner of the image.

New Mexico State University, through grant funding, studied how to
accomplish calibration of CCD-based video meteor cameras used in SkySentinel  \citep{Bannister2013}. Bannister made several simplifications to the Borovi{\v c}ka model, one of which was to replace the exponential model of zenith distance by a quadratic approximation. Applied to actual SkySentinel sensors, Bannister showed that his 8-parameter all-sky calibration model could achieve 0.21 degree accuracy in azimuth and 0.09 degree in zenith angle. For a SkySentinel sensor, this is sub-pixel precision because 1 pixel subtends about 0.3 degree on the sky.

The calibration of SkySentinel sensors is based upon Bannister's paper, but with two significant changes. First, an exponential model replaced the quadratic model of zenith distance. This is because the quadratic model of zenith distance was leading to systematic error at zenith distances greater than 70$\degree$. Most bolides are seen at large zenith distances (i.e., low elevations), and utmost accuracy is needed in this regime. Indeed, replacement of the quadratic model by an exponential model of zenith distance was a step back to the original Borovi{\v c}ka model (\citet{Borovicka1992}; \citet{borovicka1995}).

\begin{figure}
	\includegraphics[width=\columnwidth]{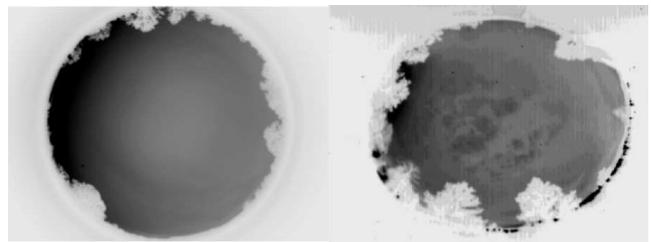}
	\caption{Image flattening. The left frame shows a flattening of $\Phi=0.0$, while the right frame shows $\Phi=0.09$} 
	\label{fig:calibration}
\end{figure}

Second, we discovered that the USB video capture devices used to digitize
SkySentinel's analog signal often introduced elliptical distortion. Figure~\ref{fig:calibration} illustrates the difference between a normal image and one that has elliptical distortion. The axes of the ellipse always appeared to align with the $x$ and $y$ axes of the sensor. The distortion was leading, in some cases, to several degrees periodic error in the residuals. To restore image circularity, $x$ and $y$ coordinates were adjusted using appropriate scale factors. Nodes 5 and 6 had measurable amounts of image ellipticity while Nodes 37 and 79 had virtually none. Table~\ref{tab:calibration} tabulates the parameters used to calibrate each node.

\subsection{Calibration Procedure}
\label{sec:calproc}
Calibration points (stars) brighter than visual magnitude +2 were selected from a list downloaded from the U.S. Naval Observatory's Online Astronomical Almanac\footnote{\texttt{http://asa.usno.navy.mil/SecH/BrightStarsSearch.html}, accessed July 01, 2015}. Hourly SkySentinel composite images were selected from a period of 4-5 days centered on the 2 June 2016 bolide event. Images were converted to TIF format, 8-bit depth, and enhanced by subtraction of a median combined image to enhance star visibility. Calibration stars were selected over a wide range of azimuths and zenith distances, limited only by availability of stars bright enough to be seen by SkySentinel. The 15-degree/hour change of hour angle made it possible to measure the same star multiple times during the course of a night.

Mira Pro x64 Windows astronomical software \citep{Mira64}
was used to measure calibration stars until there were 50-150 data points. Mira Pro defines the upper left (Northeast) pixel to have coordinates (1,1), which differs from SkySentinel's convention that this same pixel has coordinates (0,0). Accordingly, Mira Pro coordinates were decremented by 1, along each axis, before data entry into the SkySentinel calibration model. 

A Microsoft Excel 2011 
workbook was created to solve for the 10 calibration parameters using a least-squares procedure. For each calibration point, the great circle error between cataloged (USNO Bright Stars) and calculated (Calibration Model) star coordinates was calculated. The errors were squared and summed to yield a sum of squared errors (SSE). 
Excel's non-linear solver was then used to minimize SSE by varying the 10 calibration parameters, until the solver converged on a solution. Of these 10 parameters, 9 were independent, because the y scale factor was constrained to equal the reciprocal of the x scale factor. This constraint ensured that the area of the image projection remained constant.
Table ~\ref{tab:calibration} lists the values of the calibration parameters found for each of the three SkySentinel nodes that were used for the analysis of the event.

\begin{table}
	\centering
	\caption{Calibration Parameters of three SkySentinel Nodes that were used for analysis}
	\label{tab:calibration}
	\begin{tabular}{ccccc}
		\hline
		\multicolumn{1}{c}{} & \multicolumn{3}{c}{SkySentinel Node}\\
		\hline
		& 5          & 6          & 37         & 79         \\
		\hline				
		\hline
		V & 0.005891   & 0.00589    & 0.005792   & 0.005934   \\
		\hline
		S & 0.002081   & 0.001458   & 0.001791   & 0.001979   \\
		\hline
		D & 0.017449   & 0.0191     & 0.017863   & 0.017886   \\
		\hline
		a$_0$  [deg] & -22.349    & -1.14      & -2.116     & -17.739    \\
		\hline
		E [deg] & 240.671    & 59.909     & 223.085    & 289.919    \\
		\hline
		$\epsilon$ [deg] & 2.799      & 2.588      & 1.449      & 0.954      \\
		\hline
		X$_0$ & 225.405159 & 227.54618  & 235.454533 & 237.610721 \\
		\hline
		Y$_0$ & 338.665956 & 328.936475 & 320.828751 & 313.380068 \\
		\hline
		$x$ scale factor & 0.988604   & 0.988506   & 1.000282   & 0.999640    \\
		\hline
		$y$ scale factor & 1.011527   & 1.011627   & 0.999718   & 1.000360   \\
		\hline
	\end{tabular}
\end{table}

Standard deviation of the great circle error,  $\sigma = \sqrt{SSE/(n-10)}$, where $n$ is the number of calibration stars, ranged from 0.07-0.16 degree. Goodness-of-fit was verified by examining residual plots to ensure they were randomly distributed with no patterns or trends. The mean absolute deviation of zenith distance $|\delta z|$ ranged from 0.04-0.09 degree, and the mean absolute deviation of azimuth $|\delta a \sin z|$  ranged from 0.03-0.07 degree. These uncertainties, no greater than one-third of a pixel, show that SkySentinel cameras are capable of accurate calibration. 

In summary, the 10-parameter calibration model produces accurate horizon coordinates $(z,a)$ where $z$ is the zenith distance and $a$ is the azimuth measured from cardinal South. To get azimuth measured from cardinal North, compute $a + \pi$, and subtract multiples of 2$\pi$, as needed, to get a result on the interval $[0, 2\pi)$.

\section{Results From SkySentinel Data}

\subsection{Fireball Trajectory}

For the trajectory analysis of the bolide we used data from three sites: Flagstaff (Node 6), Payson (Node 79), both located in Arizona, and Albuquerque (Node 5), in New Mexico. 
We could not use data from the other cameras (Nodes 1, 2 and 14), as the bolide disappeared below the horizon or behind obstructing objects causing only the bloom to be seen in the videos. In the case of Node 37, there was an unknown debris covering the camera near the location of the bolide. This led to an inaccurate centroid for the bolide on the frame, which skewed the results for the trajectory from this camera. By carrying out sensitivity tests, later we found that increasing the calculated elevation of the bolide by about $0.2\degree$ fixed the offset, but due to the nature of the issue, it is not possible to precisely determine this value. As such, we have also excluded Node 37 data from the analysis.

The Flagstaff station was the first to detect light from this event and in this paper we use the first detection of the bolide from that video as a reference time. The MultiSpectral Radiometer (MSR) recorded data at 10,000 frames per second and used GPS clocks for timing. Consequently, we used the MSR to calibrate the timing in the other observations by aligning the features from the respective light curves. The offsets in the Node 79, Node 6 and Node 37 data were $t=0.225$ s, $t=0.2$ s and $t=0.42$ s respectively. Hereafter, all times are quoted with reference to the start of the event at 10h 56m 27.489s UTC.

We used frames from the first few seconds of the recordings up until the saturation of the pixels made it impossible to accurately determine the location of the bolide from the camera frames.  This gave us 74 frames from Nodes 6 and 57 frames from Node 79. For Node 5, we were able to use frames after $t\sim1.5$s -- i.e. 46 frames. Therefore, the first $1.5$s of the trajectory was determined only using Nodes 6 and 79, while the later part was determined using all three cameras, until Node 79 saturates. The last $0.2$s of the triangulated part of the trajectory are determined using Nodes 5 and 6.

In order to reconstruct the trajectory we turned to the plane intersection method detailed in \citet{Ceplecha1987}. Due to the locations of our cameras relative to the direction of motion of the object, the angular separation between the intersecting planes from two of the cameras (Nodes 6 and 79) was small enough ($\sim 16.7\degree$) to result in loss of their statistical significance. This made the plane intersection method impractical for the analysis, as the statistical weight of this pair is only about $5\%$. For all three cameras, however, we were able to use an alternate approach that employs the least-squares technique, similar to that described by \citet{borovicka1990}.

Using this triangulation method, for each time step we define a vector that points to the location of the bolide in a geocentric reference frame. Here, $\mathbf{R_i}$ is the unit vector pointing to the observation site from the center of the Earth and $\mathbf{r_i}$ is the unit vector pointing to the bolide from the observation site. The location of the bolide is then the vector sum with $R_i$ being the distance from the observation site to the center of the Earth and $r_i$ being the free parameter which is the distance between the observation site and the bolide. The location of the bolide in terms of the site $i$ is then:
\begin{equation}
\mathbf{h}_i = r_i\mathbf{r_i} + R_i \mathbf{R_i} 
\end{equation}

The solution to the triangulation is obtained by finding $r_i$ and $r_j$ which minimizes the distance between the position of the bolide from any pair of observations $i,j$, using the Gauss-Newton method. We define the difference vector $\mathbf{d}$:
\begin{equation}
\mathbf{d}(r_i, r_j) = \mathbf{h_i} - \mathbf{h_j} = r_i\mathbf{r_i} + R_i \mathbf{R_i} - r_j\mathbf{r_j} - R_j \mathbf{R_j} 
\end{equation}

Taking $u$, $v$ and $w$ to be the x-, y- and z- coordinates of $\mathbf{d}$, we can create the Jacobian $J$ and $b$, given by. 

\begin{equation}
J = \begin{bmatrix}
\frac{du}{dr_i} & \frac{du}{dr_j} \\[0.5em]
\frac{dv}{dr_i} & \frac{dv}{dr_j} \\[0.5em]
\frac{dw}{dr_i} & \frac{dw}{dr_j}
\end{bmatrix}
\quad
b = -\begin{bmatrix}
u(r_i, r_j) \\[0.5em]
v(r_i, r_j) \\[0.5em]
w(r_i, r_j)
\end{bmatrix}
\end{equation}

Choosing an initial guess for $\mathbf{p} = (r_i, r_j)$, the solution is found iteratively by solving for the step given by $d\mathbf{p} = (J^{T} J)^{-1}J^{T}b$, until the $|\mathbf{d}|$ is below a threshold, or $|d\mathbf{p}| = 0$. The pairs of observation each produce a solution to the meteor position as a function of time, i.e. in our case, for 3 observations, we have 6 solutions. A standard linear regression is applied along each direction for all solutions, and the corresponding residuals are determined.


The details of the trajectory parameters are listed in Table~\ref{tab:traj}.  The luminous flight began at coordinates $\phi = 34.555 \pm 0.002\degree$N planetographic latitude, $\lambda = 110.459 \pm 0.002\degree$W longitude, and a height of $100.2 \pm 0.4$ km.  

To obtain the radiant, we corrected the velocity for Earth's rotation and the curvature of the trajectory due to Earth's gravity, following the method described in \citet{Ceplecha1987}. This was then projected onto the celestial sphere defined by the geocentric inertial frame.  
The geocentric velocity was determined at the entry location using
\begin{equation}
    v_g^2  = v_{\infty}^2 - \dfrac{2GM_{\earth}}{R_{\earth} + h_b},
\end{equation}
where $M_{\earth}$ and $R_{\earth}$ is the mass of the Earth and the WGS84 radius at the current location respectively, and $h_b$ is the height of the bolide at the entry location. $v_{\infty}$ is the pre-atmospheric entry speed in the geocentric inertial reference frame. 

We also obtained the AZ/EL of the bolide radiant in the co-rotating geocentric frame at the entry point of the bolide using the AstroPy module in Python \citep{AstropyPaper}. The Arizona fireball moved closely from North to South, with the azimuth of the radiant being $13 \pm 1 \degree$, measured from North. 

From the SkySentinel observations we could not determine the altitude of the maximum brightness or the terminal height of the luminous flight due to image saturation and the obscured view that we described above. 

\begin{figure}
	\includegraphics[width=\columnwidth]{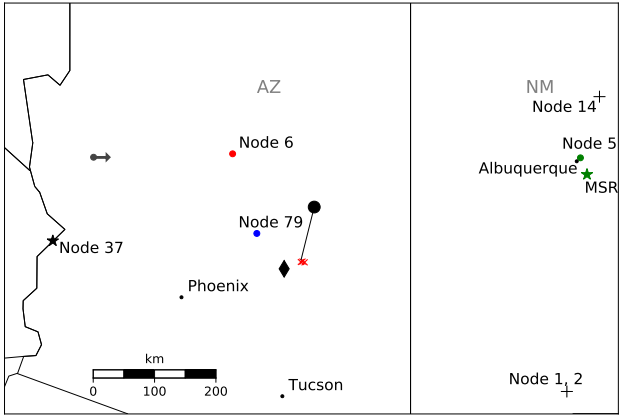}
	\caption{The trajectory of the bolide with the entry location marked with a black circle. Coloured circles indicate the cameras that we used to determine the trajectory. The stars represent the cameras that were only used for light curve analysis, while the black crosses denote cameras that observed the bolide but were not used in the calculations. Nearby cities are marked with the small black circles. Locations of the fragments found are marked by the red $\times$'s. The gray circle is the location of the car whose dash cam video was used in the analysis. The arrow indicates the direction that the car was travelling. The black diamond shows the peak brightness location from CNEOS database.}
	\label{fig:trajectory}
\end{figure}

\begin{figure}
	\includegraphics[width=\columnwidth]{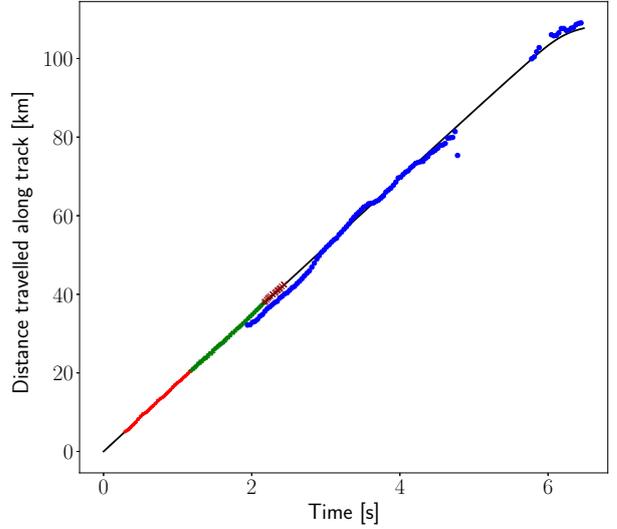}
	\caption{The distance travelled by the bolide as a function of time. The red points near the start of the event are from the triangulation of data SkySentinel Nodes 6 and  79, the green crosses ($+$) are using all three cameras, the dark-red $\times$'s are using Nodes 5 and 6, while the blue points are from the dash cam video. The solid line is from the fragmentation model. }
	\label{fig:Lt}
\end{figure}

\begin{figure}
	\includegraphics[width=\columnwidth]{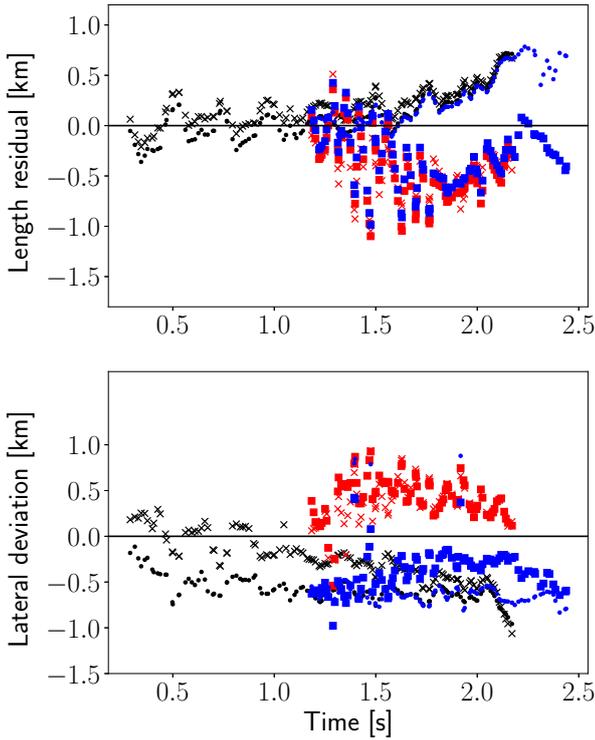}
	\caption{The residual in the expected length along track (top), and the lateral deviation from the average trajectory (bottom) from the triangulation solution using SkySentinel cameras. Node 6 are plotted with $\times$'s, Node 79 with points and Node 5 with squares. The colors represent different pairs used for the solution: black for Nodes 6 and 79, red for Nodes 6 and 5, and blue for Nodes 79 and 5. }
	\label{fig:residual}
\end{figure}

\begin{figure}
	\includegraphics[width=\columnwidth]{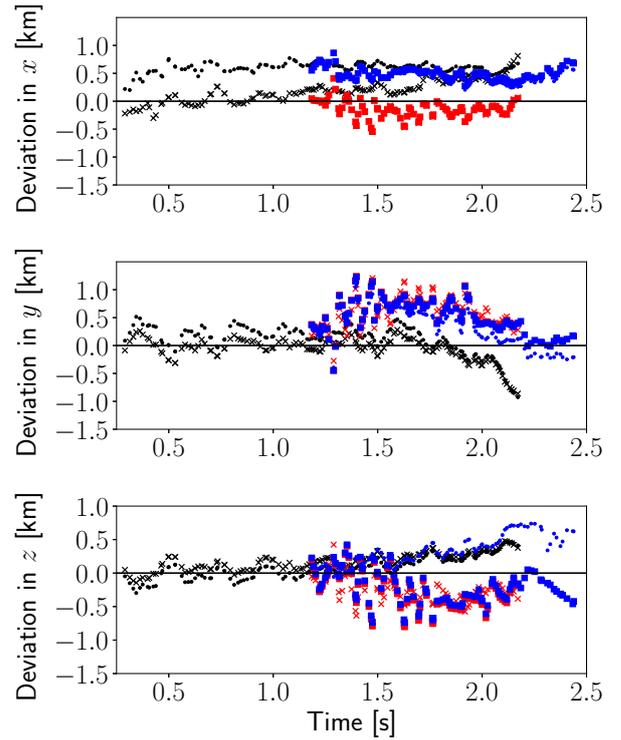}
	\caption{The residuals along the $x$, $y$ and $z$-direction from the linear fit. The color coding and markers follow the scheme given in Fig~\ref{fig:residual}. }
	\label{fig:xyzdiff}
\end{figure}

Figure~\ref{fig:trajectory} illustrates the trajectory, along with a map of the nodes which observed the event. Figure~\ref{fig:Lt} shows the distance travelled by the bolide along the track as a function of time while Figure~\ref{fig:residual} and~\ref{fig:xyzdiff} show the scatter in the results. Node 5 has the largest scatter, although this is likely due to it being much further away compared to the other nodes. The average deviation is about $200$m.

\begin{table*}
	\centering
	\caption{Trajectory and orbital parameters using triangulation of the bolide's position from the initial frames of SkySentinel cameras, the analysis of the dash cam video and the fragmentation model. Azimuth and zenith distance are given for the apparent radiant in the co-rotating geocentric frame. The location of the geocentric radiant is given on the celestial sphere with corrections applied for rotation of the Earth and curvature of the trajectory by Earth's gravity. }
	\label{tab:traj}
	\begin{tabular}{lcp{6em}lc}
		\hline
		\multicolumn{2}{c}{Trajectory}	& & \multicolumn{2}{c}{Orbit} \\
		\hline
		\hline
		Entry Latitude$^*$ 							&  34.555 $\pm$ 0.002$\degree$ N &  ~	& Semi-major axis 				& 1.13 $\pm$ 0.02 AU \\
		Entry Longitude$^*$ 						& 110.459 $\pm$ 0.002$\degree$ W &	~	& Eccentricity 						& 0.210 $\pm$ 0.01 \\
		Entry Height$^*$ 							& 100.2 $\pm$ 0.3 km			 &	~	& Perihelion distance 		& 0.89 $\pm$ 0.01 AU \\
		Pre-atmospheric Velocity$^*$ 		        & 17.4 $\pm$ 0.3 km/s			 &	~	& Inclination  						& 23.2 $\pm$ 0.7$\degree$ \\
		Peak brightness magnitude$^\dagger$ 		&	-20.4 $\pm$ 0.2 			 &	~	& Ascending node angle 		& 72.271 $\pm$ 0.001$\degree$ \\
		Height of peak brightness$^\dagger$ 		& 29.8 $\pm$ 0.6 km				 &	~	& Argument of perihelion 	& 109 $\pm$ 6$\degree$ \\
		End height $^\dagger$						& 21.9 $\pm$ 0.6	km			 &	~	&  & \\
        End latitude $^\dagger$						& $33.924\degree$ N  			 &	~	& \multicolumn{2}{c}{Apparent radiant at entry location} \\
        End longitude $^\dagger$					& $110.641\degree$ W			 &	~	& Zenith distance of radiant$^*$				& 42 $\pm$ 1$\degree$ \\
		                                            &                    			 &	~	& Azimuth of radiant & $13 \pm 1 \degree$ \\
		                                            &                                &  ~   & \multicolumn{2}{c}{Geocentric radiant (J2016.5)} \\
		                                            &                                &  ~   & Right ascension & $349 \pm 3 \degree$   \\ 
    	 	                                        &				                 &  ~   & Declination& $78 \pm 1 \degree$ \\ 
    	 	                                        &                                &  ~   & Geocentric velocity & $13.3 \pm 0.3$ km/s \\ 
		\hline
		\multicolumn{3}{l}{$^*$ from triangulation using SkySentinel cameras} & & \\*
		\multicolumn{3}{l}{$^\dagger$ using data from dash cam video and the fragmentation model.}\\*
	\end{tabular}
\end{table*}

\subsection{Velocity}
\label{sec:velocity}

Performing a linear fit to positions retrieved by triangulating SkySentinel camera data between 0.4s and 2.5s 
gives an observed velocity of $17.4 \pm 0.3$ km/s with respect to the rotating surface of the Earth. 
During this time interval, 
there is no observed deceleration of the bolide. 
Thus, we consider this value as the pre-atmospheric entry velocity that remains constant until the initial breakup.

We could not include data for the later stages; during the main flare event the frames were saturated and after de-saturation, the bolide disappeared behind surface objects, e.g., a tree in the case of Node 79. This made it impossible to determine from SkySentinel data, the change in velocity during and after the breakup of the object. 

\subsection{Light curve}

For the lightcurve analysis, we attempted photometric calibration of the SkySentinel cameras. These datasets were from 8-bit grayscale cameras in the visible wavelength, observing at 30 frames per second. The bolide was below the horizonal throughout the entire video for Nodes 1, 2, and 14, while for Node 5, the bolide was visible only for the first two seconds. Therefore, only three cameras (Nodes 6, 37 and 79) observed the entire event. 



\begin{figure}
	\includegraphics[width=\columnwidth]{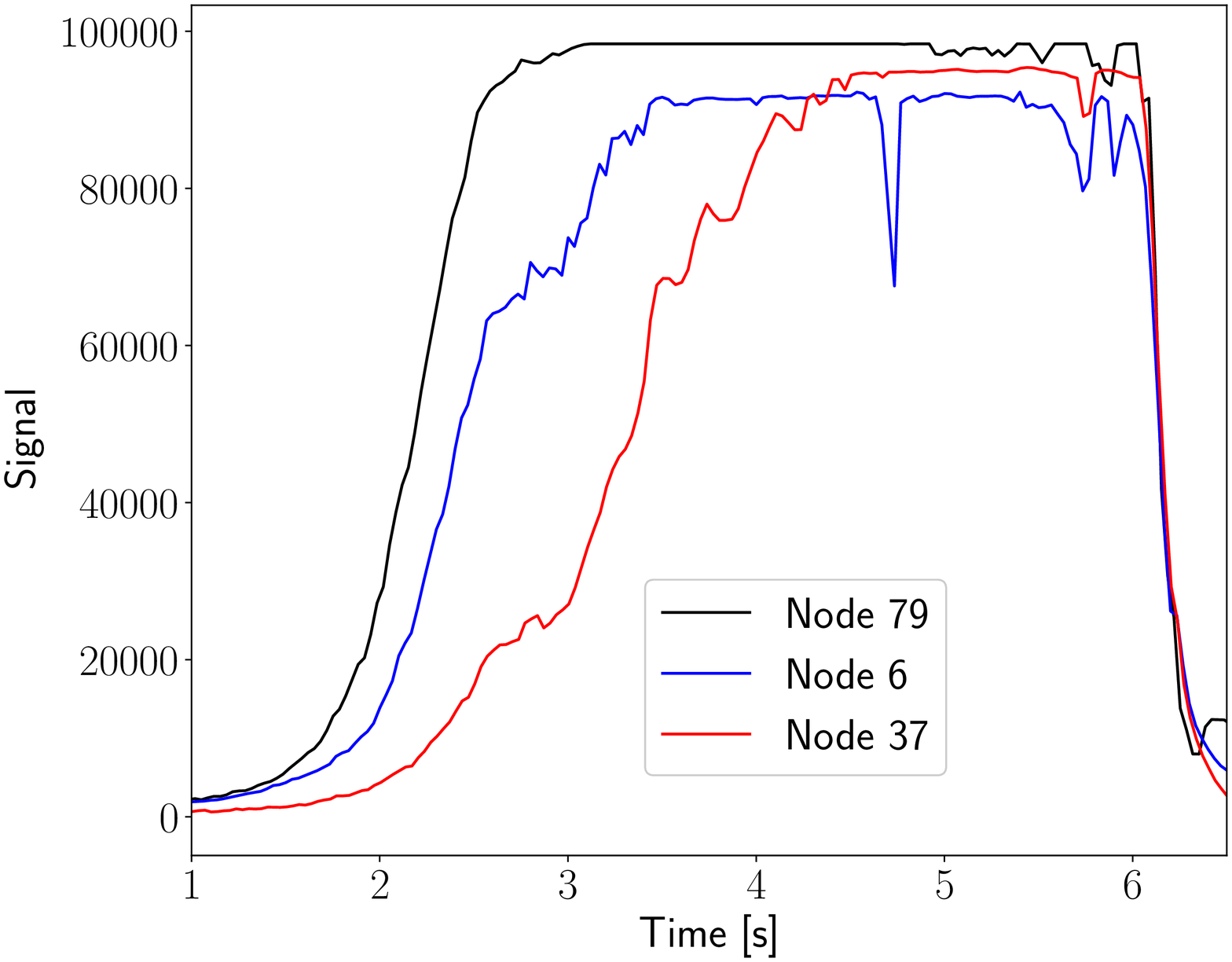}
    \caption{The signal from the SkySentinel cameras in arbitrary units. All three cameras were saturated during the main flare event, making the data unusable.}
    \label{fig:Mnodes}
\end{figure}

The sum of the raw pixel counts were taken to  be  proportional  to  the  intensity  of  the  light  observed, and thus taking the base-10 logarithm and multiplying by a factor of -2.5 provided an uncalibrated instrumental magnitude. A photometric calibration of the SkySentinel was attempted using the Moon and Vega. An image of the Moon was used on a later date for all four cameras. However, the Moon's light saturated the camera, making accurate calibration difficult. In the case of nodes 6 and 37, Vega was visible, but since the SNR of Vega was $\sim 3$ in both cameras, we were unable to reliably determine the photometric correction. Furthermore, the raw signal from three cameras (Nodes 6, 37 and 79) are clearly saturated (Fig~\ref{fig:Mnodes}), so the SkySentinel cameras were not used in the photometry analysis.

\begin{figure}
	\includegraphics[width=\columnwidth]{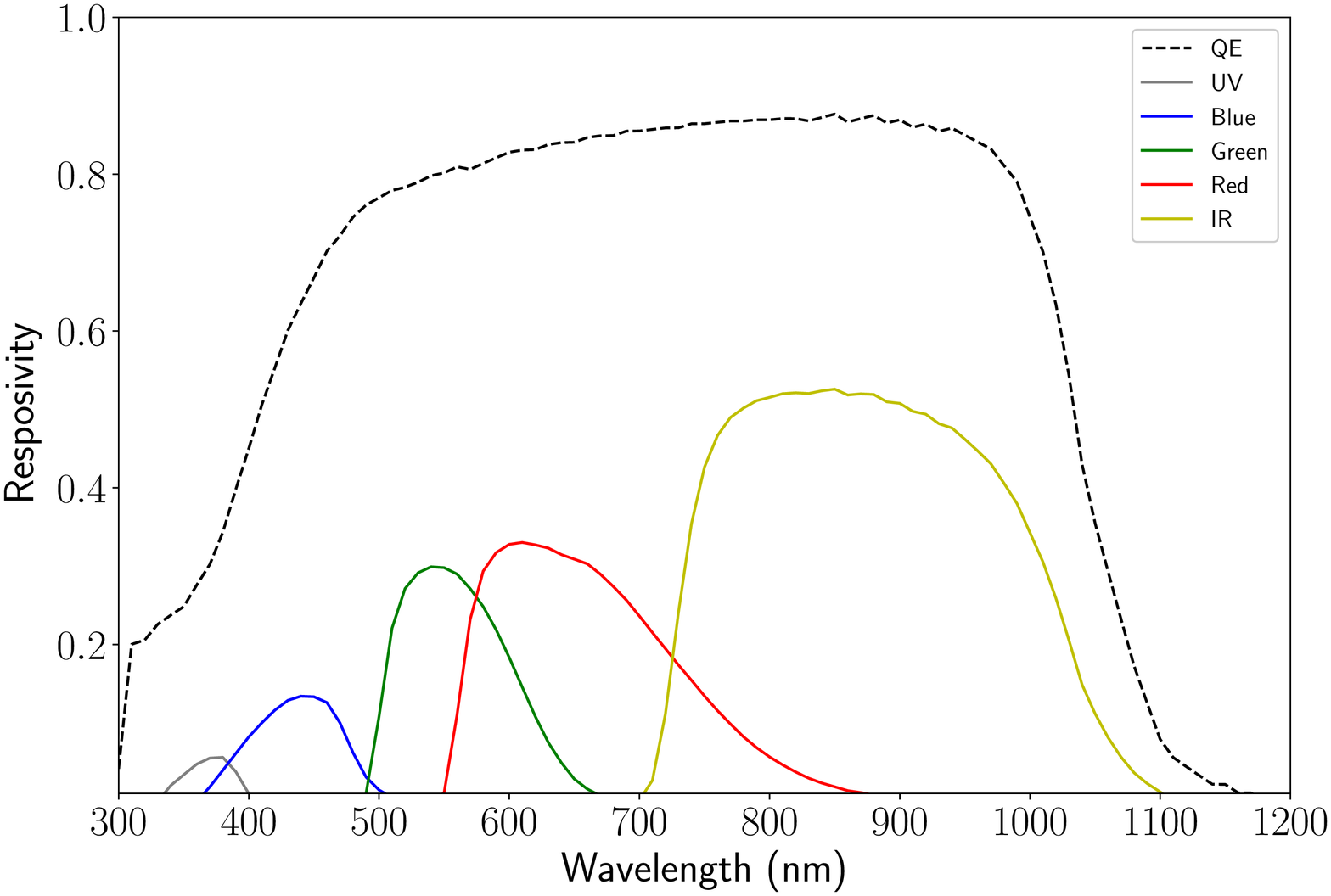}
	\caption{The dashed curve shows the quantum efficiency of the detector, while the solid lines are the responsivity (A/W) of each filter - from left to right: UV, blue, green, red, IR }
	\label{fig:MSRQE}
\end{figure}

\begin{figure}
	\includegraphics[width=\columnwidth]{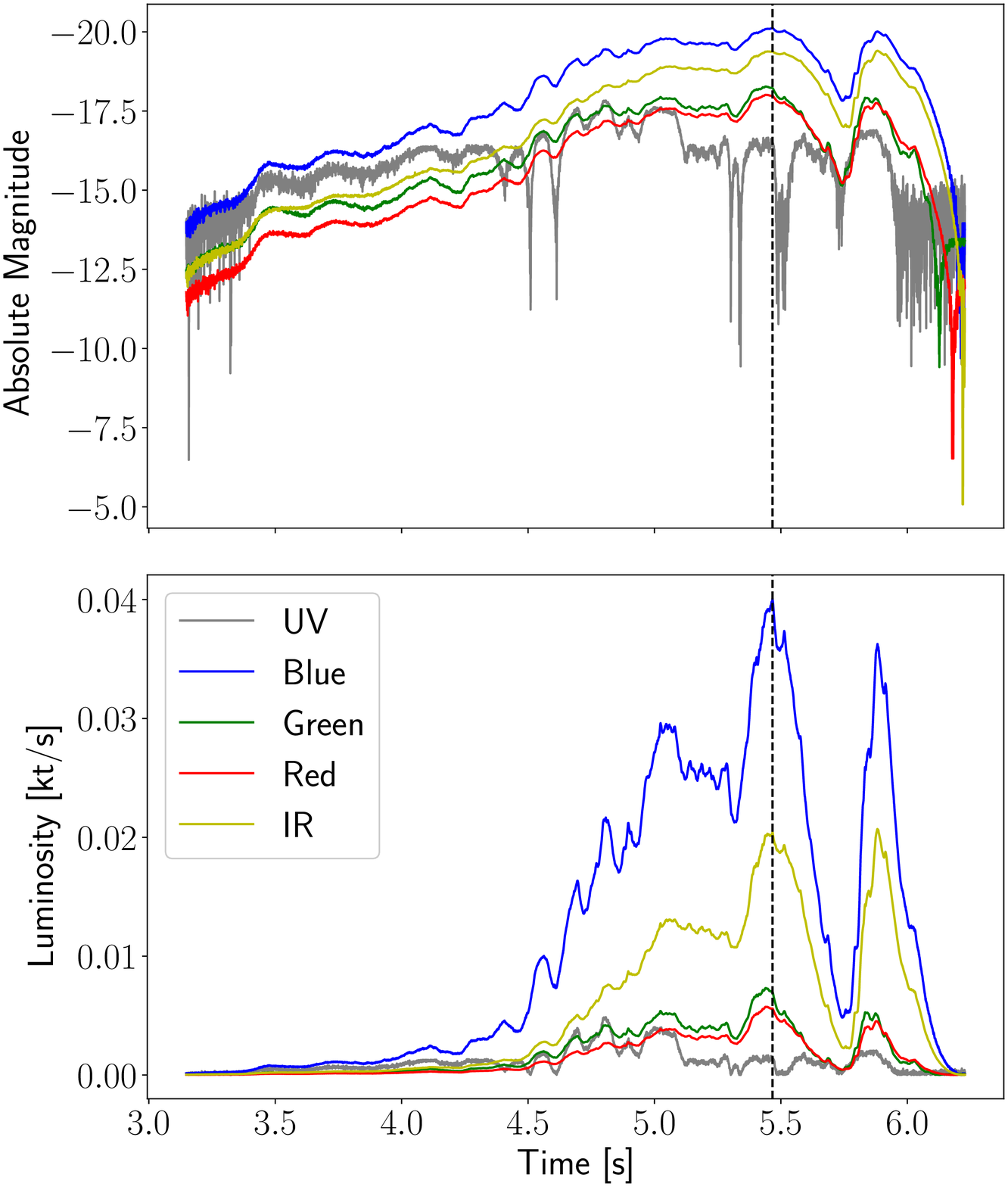}
	\caption{The absolute magnitude and luminous power of the bolide from the different filters of the MSR as a function of time. The vertical dashed line indicates the peak brightness.}
	\label{fig:MLMSR}
\end{figure}

We also used data from the MSR instrument in New Mexico, which observed at 5 bands: UV, Blue, Green, Red and Infrared at 10,000 frames per second. The quantum efficiency and responsivity of the MSR were already known (Figure~\ref{fig:MSRQE}) and thus, it was a straightforward matter of converting the raw signal from the MSR into a calibrated magnitude, from which we obtained the luminosity. We take a power of $1500$ W at $100$km to correspond to an absolute magnitude of of zero \citep{Ceplecha1998}. The magnitude values were corrected for extinction at each time stamp, with the extinction coefficients determined from calibrating an image of the Moon at different airmass. Figure~\ref{fig:MLMSR} plots the absolute magnitudes and luminosity from the MSR data.

After the initial brightening, the first notable flash occurs at around 4s, followed by multiple bright flaring events. There are two distinct peaks in the light curve, at around 5.5 s and at 5.9 s. 

We determined the total luminous energy of the fireball by integrating the light curve. We use only the visual magnitudes (B, V and R). The corresponding total impact energy ($E$) was calculated based on the total luminous energy ($E_0$) using the empirical relation by \citet{brown2002nature}:
\begin{equation}
E = 8.2508\times(E_0)^{0.885}
\end{equation}
The impact energy then can be used, along with the pre-atmospheric speed to calculate the mass of the object. The results are shown in Table~\ref{tab:light}. The calculated average energy of the incoming bolide was about $0.54\pm0.06$ kt which yields an estimated mass of $14.8\pm1.7$ metric tonnes.

\begin{table}
	\centering
	\caption{Results of light curve analysis.}
	\label{tab:light}
	\begin{tabular}{ccc}
		\hline
		& & MSR   \\
		\hline
		\hline
		\multirow{2}{*}{$E_0$} 
		& [$\times 10^{12}$ J] 	& 0.191  \\
		& [kt]  & 0.0455 \\
		\hline
		\multirow{2}{*}{$E$} 
		& [$\times 10^{12}$ J]  & 2.24 \\
		& [kt]  & 0.536 \\
		\hline
		\multicolumn{2}{c}{Mass [metric tonne]}  & 14.8 \\
		\hline
	\end{tabular}
\end{table}

\subsection{Orbit}
We determined the pre-atmospheric orbit from the velocity values measured at the earliest part of the bolide's trajectory. First, we calculated the heliocentric position and velocity vectors at the entry point. Then we carried out a backward time-integration using the REBOUND code \citep{rein2011} to determine the position and velocity outside the Hill sphere of the Earth. We tested several integration times to ensure consistency in the retrieved orbit. There were small variations in the orbital parameters up to about 1.5 years before the event, after which the change was insignificant. Therefore, we integrated for 1.5 years, at which point we obtained the heliocentric orbital parameters. These values are listed in Table~\ref{tab:traj}. 

\begin{figure}
	\includegraphics[width=\linewidth]{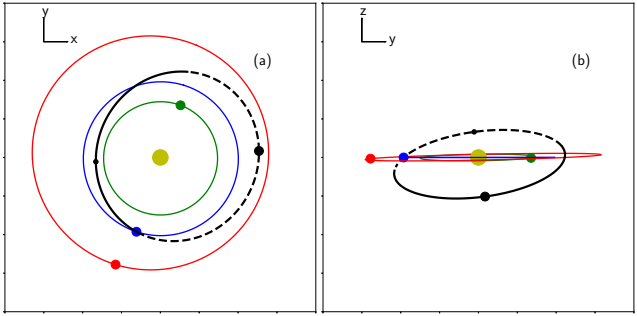}
	\caption{The diagram of the calculated orbit is shown in black, with the solid line above the ecliptic and dashed below in (a). (b) shows the view down the x-axis with the solid line before the y-z plane and dashed line, behind. The small and large black points show the peri- and aphelion respectively. The orbits of Venus, Earth and Mars are shown along with their positions on June 2, 2016. The vernal equinox is to the right in (a) and out of the page in (b). }
	\label{fig:orbit}
\end{figure}

\begin{figure}
	\includegraphics[width=\columnwidth]{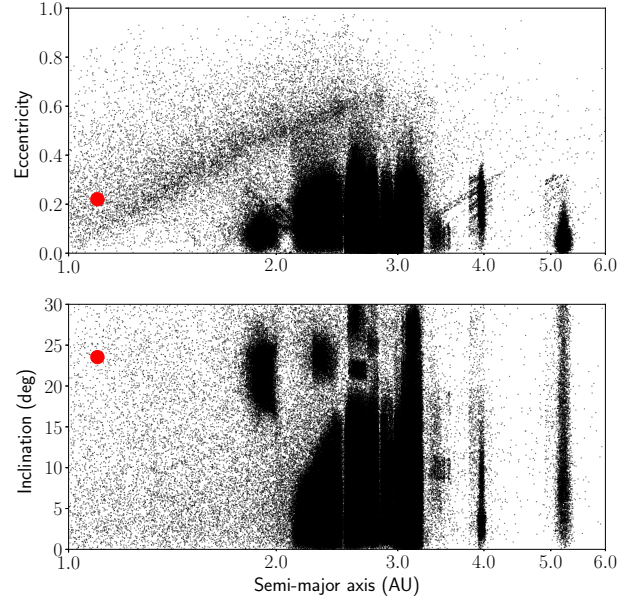}
	\caption{The asteroidal and cometary orbital element distribution of known objects are shown in small black dots. The orbital element of the bolide is plotted with the large red circle. There are no distinguishable families that the bolide forms part of.}
	\label{fig:orbitalelem}
\end{figure}

Figure~\ref{fig:orbit} illustrates the orbit of the impactor which was inclined and slightly eccentric. The object reached perihelion about 64 days before the impact event. Figure~\ref{fig:orbitalelem} shows the bolide's orbital parameters plotted against the orbital element distributions of known asteroids and comets. The object clearly did not belong to any known main-belt asteroidal families, but was part of the Apollo category of near-Earth objects.

A search for the parent body was done using the Drummond criterion \citep{Drummond1981} and the modification by Jopek \citep{Jopek1993} and data from the MPCORB\footnote{{\texttt https://www.minorplanetcenter.net/iau/MPCORB.html} accessed December 2, 2017} database. This yielded no match for any known parent body. 

\section{Additional Observation and Modeling}

A significant number of movies became available on video sharing websites and news portals that were recorded by dash cams and security cameras.  Most of these have the problem of obstructed view throughout most or all of the bolide's flight path and show only lens flare. In a few instances the entire event was clearly visible and for one of these videos we were able to identify the exact location where it was recorded from. We have analyzed this video to give us additional information on the bolide properties. Furthermore, we have set up a fragmentation model to reveal pieces of the puzzle that was unavailable to us from the observational data only. 

\subsection{Dash Cam Footage}
Mark Olvaha recorded the event using his GoPro dash cam\footnote{\url{https://www.youtube.com/watch?v=o-KED81DO5c}, accessed August 18, 2017}.  Based on the land and road features seen in the video and in the video description, we were able to pinpoint the location of the car at the time of the recording as being on I-40 east of Kingsman, AZ at ($35.159$ N, $113.687$ W).  For our analysis we used a revised version of this video\footnote{\url{https://www.youtube.com/watch?v=hDoUQnxY7z4} - 0:00 to 0:31, accessed August 18, 2017} because in this version the authors applied image stabilization to the original source that made the bolide tracking easier. Since the position where the dash cam movie was recorded was significantly further away compared to the SkySentinel camera locations ($\sim 350$km), we were only able to get the light curve and the track of the object close to the end of the observation (after $t\sim 3$s). We calibrated the footage timing with the MSR dataset by matching the peaks of the sum pixel values from the video.

\begin{figure*}
	\includegraphics[width=\textwidth]{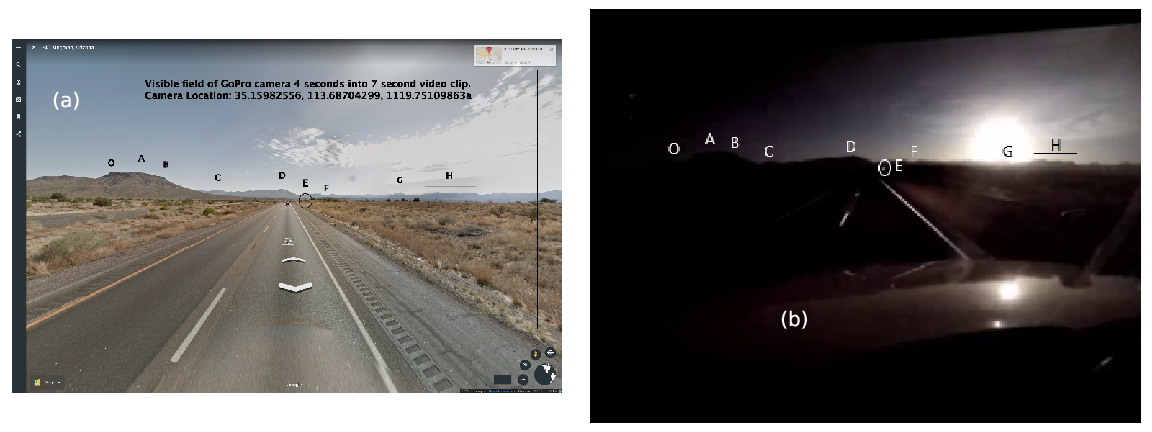}
    \caption{Angular calibration of dashcam video using surface features.  }
    \label{fig:dashcamcalib}
\end{figure*}

Similar to the method described in section~\ref{sec:calproc}, the video was calibrated for astrometry using local geographic features as shown in Fig~\ref{fig:dashcamcalib}. 
Features O, B, C, D, F, and G (six topographic points total) were used. Features A, E, H were additional markers that potentially could have been used for calibration but were not. A is a radio antenna that was not visible in the nighttime video; E is an interstate highway sign whose azimuth changed significantly during the course of the video; H is a mountain range without a well-defined summit. The topographic features had well-defined silhouettes that facilitated determination of their azimuths and apparent elevations relative to the local horizon of the car. Using Google Earth, it was straightforward to obtain azimuth from the known position of the car to each topographic feature. 
To calculate the apparent elevation of each topographic feature above the car's horizon, a flat earth model was assumed. Each feature's elevation above the car's horizon was obtained by first computing the difference in altitude (expressed in terms of distance above mean sea level) between the car and the feature's summit. Each feature's apparent elevation, $el = \arctan (a/d)$ was calculated, where $a$ = altitude difference and $d$ = distance to the feature. Google Earth's topographic data and measurement tools were essential to this work.

To generate an azimuth calibration model, the measured x-coordinates of the six topographic features seen on the stabilized video image were then regressed against their azimuths. This resulted in an azimuth calibration model with R-sq (adj) = 99.9\% and standard deviation = 0.98 degree. Similarly, the measured y-coordinates of the features were regressed against their apparent angular elevations. This yielded an elevation calibration model with R-sq (adj) = 98.0\% and standard deviation = 0.26 degree. Minitab statistical software was used to do the linear regressions.

The calibration models had some limitations. The apparent elevation of the highest topographic feature was 5 degrees, but the car's video imagery first saw the bolide when it was about 15 degrees above the car's local horizon. Extrapolation from 5 degrees to 15 degrees elevation necessarily introduced significant uncertainty in the bolide's elevation. At an elevation of 15 degrees, the elevation uncertainty computed by Minitab statistical software is +/- 2 degrees (95\% confidence). The uncertainty of the bolide's azimuth throughout its trajectory was also about +/- 2 degrees (95\% confidence). The AZ/EL of the features used are given in Table~\ref{tab:geofeat}.

\begin{table}
	\begin{center}
		\caption{List of geographical features used to calibrate the dashcam footage.}
		\label{tab:geofeat}
		\begin{tabular}{ccc}
			\hline
			Feature & AZ [deg] & EL [deg] \\
			\hline
			\hline
            O & 39.16 & 5.32 \\
            B & 48.81 & 5.29 \\
            C & 61.53 & 2.58 \\
            D & 80.76 & 3.09 \\
            F & 94.64 & 0.68  \\
            G & 115.55 & 1.95 \\
			\hline
		\end{tabular}
	\end{center}
\end{table}

By fixing the direction of travel (i.e. assuming that the bolide did not deviate significantly from this path), we were able to convert the angular distance on the image to a linear distance travelled by the bolide from the initial frame. 

This allowed us to constrain the distance travelled and the speed of the bolide during and after the main fragmentation, where the SkySentinel cameras were saturated or their view were obstructed. The blue points in Figure~\ref{fig:Lt} show the results of the dash cam analysis. The error in the astrometry is on the order of about $0.8\degree$ which corresponds to a error in the distance of about $5$km. Therefore, we were unable to obtain a precise deceleration profile for the bolide. However, from the last few datapoints, it is clear that the bolide has decelerated significantly. A fit of the velocity for $t\gtrsim6$s gives a velocity estimate of about $8$-$10$km/s. 


\subsection{Fragmentation Model}
To obtain information on the deceleration of the bolide as observed by the dash cam video, the position of the peak brightness and the end of the luminous phase, we implemented the fragmentation model of \citet{Borovicka2013} to calculate the post-breakup trajectory.
We assume that the only processes that causes mass loss are discrete fragmentation into one or more fragments and ablation.

While erosion and the release of dust are important to accurately calculate the mass loss, atmospheric trajectory and 
the energy deposition, we opted to exclude these processes from the model because we could not properly validate their parameterization due to lack of precise velocity data during and post saturation of the bolide.
%
%
%
To reduce the number of free parameters, we also assume that only the main body breaks into smaller fragments and that ablation is the only mass-loss process for the fragments.

In this model, we input the times of discrete gross-fragmentation events (assumed to be local maxima in the light curve), the mass loss during fragmentation and the estimated number of fragment produced. The initial parameters for the model are the entry velocity, entry position, zenith distance and azimuth of the bolide radiant obtained from the SkySentinel triangulation. 
We approximate the mass loss during each fragmentation event to be proportional to change in brightness of the respective peak in the light curve, which allowed us to distribute the mass loss among the flaring events. To be consistent with our energy analysis, we use the luminous efficiency calculated using the \citep{brown2002nature} study, which in our case is $8.52\%$. 

\begin{figure}
	\includegraphics[width=\columnwidth]{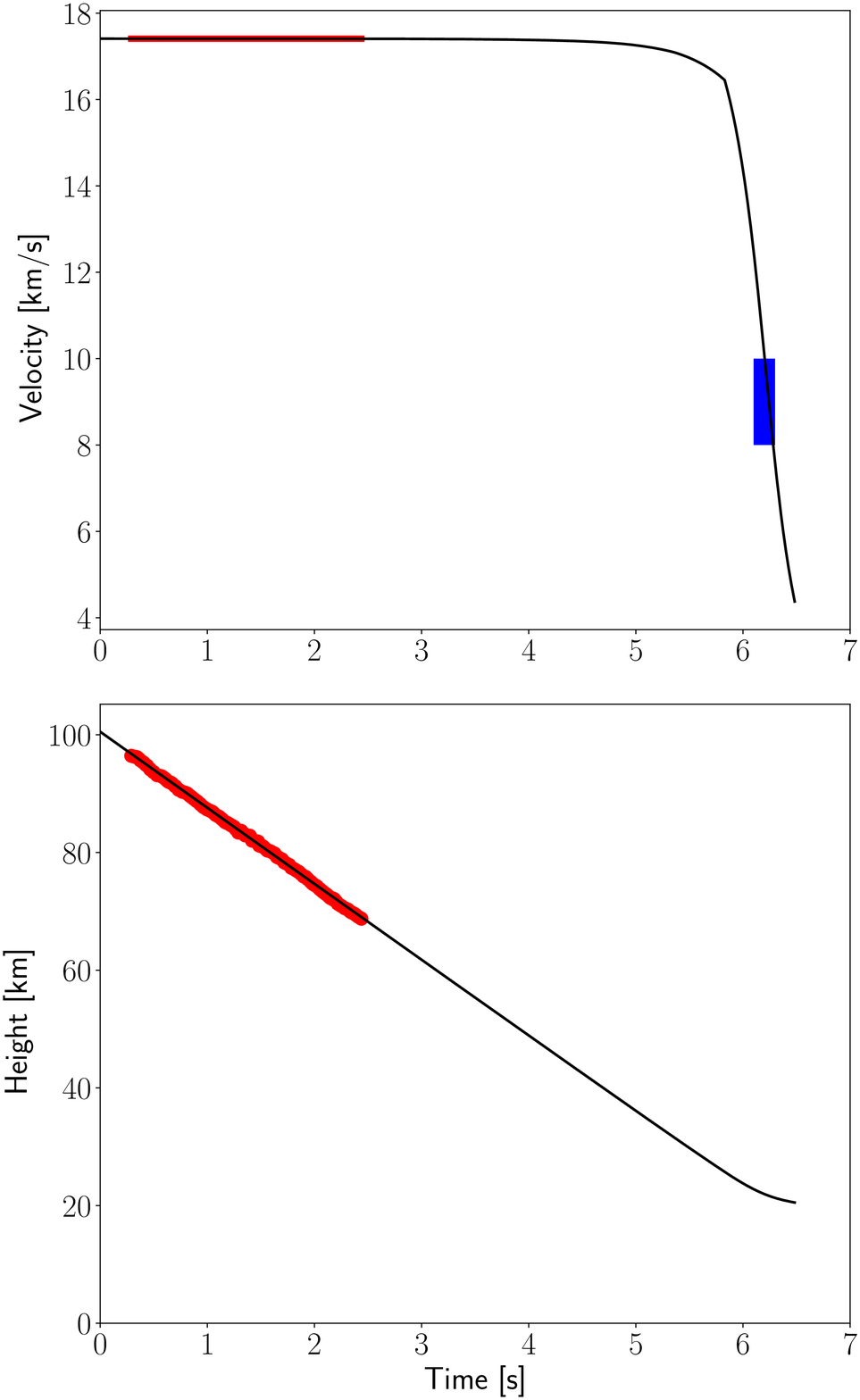}
	\caption{The modeled velocity and height of the bolide are shown as a function of time. The solid black line is from the fragmentation model using the best fit values of $K$ and $\sigma$, while the solutions of the triangulation are the red points. The solid red line in the velocity plot is the constant velocity approximation used in the triangulation from SkySentinel data. The blue box is the approximate velocity constraint of $6-8$km/s for $t$ between $6$ and $6.2$s from the dashcam video. Due to the uncertainty in the dashcam result, we were unable to determine the velocity with better precision. }
	\label{fig:VHt}
\end{figure}

\citet{Ceplecha2005} used results from observations to constrain the values of $K$ and $\sigma$. In our case, there were no observations during the latter part of event, thus we carried out sensitivity test for these unconstrained parameters. 
We varied the value of $K$ from $0.1$ to $1.0$ (c.g.s) and $\sigma$ from  $0.01$-$0.1$ s$^2$/km$^2$.

The model is validated by matching the light curve as well as the bolide positions from the SkySentinel and dashcam video. The values of $K$ and $\sigma$ are kept constant throughout the trajectory, while the mass loss and number of fragments are varied to match the light curve. The end velocity of $\sim 8$-$10$ km/s at $t\gtrsim6$s, from the dashcam video, is used a constraint to determine estimates of $K$ and $\sigma$. 

In general, for $K < 0.2$ the bolide reaches an altitude of $15$ km with a velocity greater than $12$km/s at $t\sim6.5$s, while for $K>0.6$, there is insufficient energy to produce the bright peak at $5.9$s. $\sigma\sim0.1$ s$^2$/km$^2$ causes the bolide to brighten very quickly at $h\gtrsim 70$km and mass loss to reduce the bolide into sub-kilogram fragments well above $25$km. For $\sigma<0.01$ s$^2$/km$^2$  the bolide continues to flare well after $8$s. In each of these cases, the mass loss at each fragmentation point is kept constant. While these tests by no means produce unique solutions to the fragmentation model, they are used to determine the type of object and minimize the number of free parameters to fit the observed light curve.

\begin{figure}
	\includegraphics[width=\columnwidth]{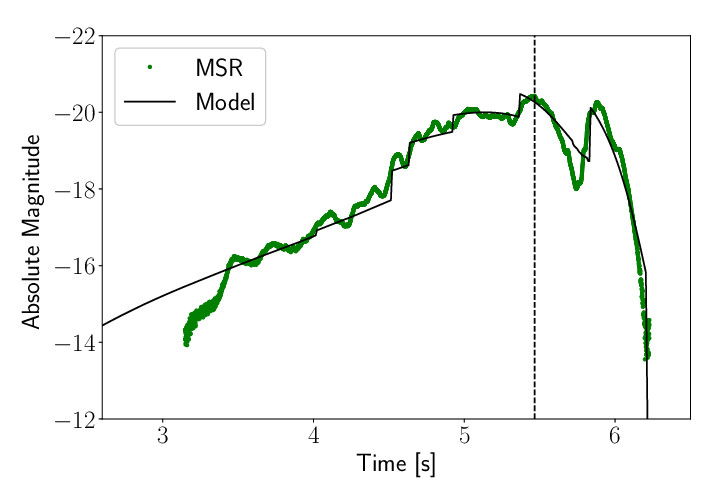}
	\caption{light curve from the model and MSR data. }
	\label{fig:fraglc}
\end{figure}

To this end, we maintain the values of $K= 0.35$ cm$^2$/g$^{2/3}$ and $\sigma= 0.048$ s$^2$/km$^2$ for the duration of the flight. 
The mass loss during each fragmentation event is shown in Table~\ref{tab:fragmass}, and resulting light curve is plotted in Fig~\ref{fig:fraglc}. The resulting total luminous energy from the model is about 2\% smaller than the luminous energy from the MSR data.

The bolide reached maximum brightness of $-20.4 \pm 0.2$ magnitudes at 5.5 s at a height of $29.8 \pm 0.6$ km. The end height of the luminous phase is at a height of $21.9 \pm 0.6$ km at $6.2$s at a planetographic latitude of $33.924 \pm 0.002 \degree$ N and longitude $110.641 \pm 0.002 \degree$ W. 

These parameters produced results that are in good agreement with both the observed light curve and the trajectory. There are, however, a few instances where the model predictions differ from observational data:

\begin{table}
	\begin{center}
		\caption{Mass loss distribution during the fragmentation events.}
		\label{tab:fragmass}
		\begin{tabular}{ccc}
			\hline
			Time [s] & Fragment mass [kg] & Number of fragments \\
			\hline
			\hline
			4.00  & 7  & 20 \\
			4.50  & 27  & 40 \\
			      & 24  & 40 \\
			4.62  & 34  & 50 \\
			      & 25  & 45 \\
			4.91  & 18  & 45 \\
			      & 24  & 35 \\
			      & 27  & 30 \\
		    5.35  & 14  & 50 \\
		          & 16  & 45 \\
		          & 8   & 60 \\
		    5.8   & 29  & 60\\
			\hline
		\end{tabular}
	\end{center}
\end{table}

\begin{enumerate}
    \item The dashcam video shows that the bolide penetrated a few kilometers deeper than the fragmentation model predicts, and the deceleration occurs at a much lower altitude. While it is possible to decrease the shape-density and ablation coefficients to match this, the resulting model bolide does not decelerate quickly enough to match the velocity at $t\sim6.2$s. Therefore, it is more likely that inconsistency in the penetration depth is an astrometric error due to the low quality of the dashcam video.
    \item The peak at $t\sim5.5$s from the fragmentation model continues to flare for more than $0.3$s, which is contradictory to the MSR data where the flare is very short-lived. This is most likely due to the fact that fragments were probably smaller than the estimated by the fragmentation model. However, decreasing the modelled mass loss produces a much dimmer peak, so either the bolide's mass or velocity are in error for the fragmentation event. Since there is no observational data for the trajectory at this location, nor are there any observed fragments in the videos, it is not possible to determine the exact set of parameters to match the light curve at this point.
    \item Initially the brightness of the model bolide is higher than observed by the MSR data.  We were able to reduce this offset by changing the ablation coefficient to $0.02$ s$^2$/km$^2$ at earlier parts of the trajectory, but it is unclear why this would change throughout the trajectory to this degree. Since we have ignored the effects of dust release and erosion throughout the flight, the contribution from these processes would possibly explain the noted deviation.
    \item The peaks at $t=5.5$s and $t=5.9$s have a smooth flare as opposed to the sharp increase caused by the gross fragmentation. These are likely eroding fragments as the mass is slowly released from erosion. However, the absence of deceleration information makes it impossible to constrain the exact mass loss from erosion and the size of grains (which determines the intensity of the peaks), so we have instead modelled these peaks as the bolide exploding into numerous, small fragments. 
\end{enumerate}



\begin{figure}
	\includegraphics[width=\columnwidth]{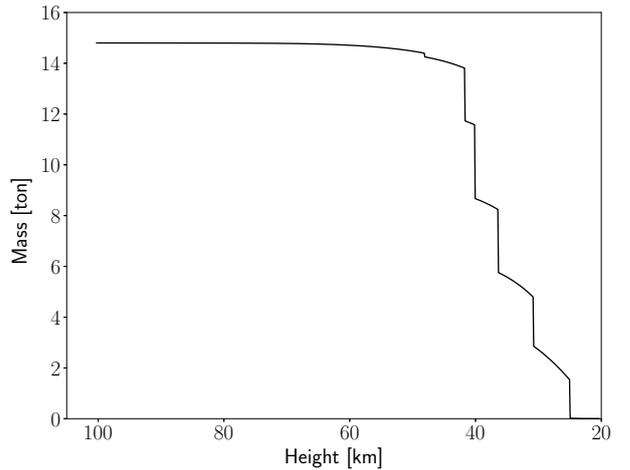}
	\caption{The mass of the main body as a function of height from the fragmentation model. The vertical drops are from discrete fragmentation events, while the smooth mass loss is from ablation. }
	\label{fig:mh}
\end{figure}

\subsection{Recovered Fragments}

Based on analysis of weather Doppler radar images, researchers from the Arizona State University's Center of Meteorite Studies found fragments from the parent body on the lands of the White Mountain Apache Tribe. 15 fusion-crusted stones were recovered during the 2 day search with a total of 79.46g of material \citep{METBulletin}. The locations of these Dishchii'bikoh meteorites are shown on Figure    \ref{fig:trajectory}. 

\begin{table*}
	\begin{center}
		\caption{Result of size estimation from different methods. The errors in the sizes are carried through from the result for the total mass.}
		\label{tab:sizes}
		\begin{tabular}{lccc}
			\hline
			Method  & Type/Material & Bulk density [kg/m$^3$] & Diameter [m]   \\
			\hline
			\hline
			PE criterion \citep{Ceplecha1976} & IIIa & 750$^\text{a}$ &  $3.4 \pm 0.1$ \\
			\hline
			\multirow{2}{*}{Peak brightness \citep{Brown2016}} & Type I & 3000$^\text{a}$ & $2.11\pm 0.08$ \\
			& Type II & 2000$^\text{a}$ & $2.42\pm 0.08$ \\
			\hline
			MSR visible bands analysis &  Fe-poor & 2500$^\text{b}$ & $2.25 \pm 0.07$ \\
			\hline
			Recovered Fragments \citep{METBulletin}
			& LL7 chondrite &  3220 $\pm$ 210$^\text{b}$& $2.04 \pm 0.11$ \\
			\hline
		\end{tabular}
	\end{center}
    \begin{flushleft}
	$^\text{a}$ \citet{Ceplecha1988}
    
    $^\text{b}$ \citet{britt2003}
    
    \end{flushleft}
\end{table*}

\subsection{Size estimation}
To calculate the size of the bolide from its mass, we require an approximation for the bulk density value.  We adapted multiple methods to determine this value indirectly from observations and we compare those results with properties of the actual ``ground truth'' value obtained from the recovered fragments.

\citet{Ceplecha1976} diagnosed the structure and bulk density values of impacting objects based on the terminal height of the fireballs they produced.  Following their method, the end height of $21$ km and entry velocity of $17.7$ km/s points to a weak cometary material (Type IIIa). For a density of $\rho \sim 750$ kg/m$^3$, we get a size of $3.3 \pm 0.1$m. However, such a weak material would probably not be able to penetrate to a depth of $\sim20$km, so this is likely an inaccurate estimate of the bolide size.

\citet{Brown2016} classified fireballs based on the analysis of their peak brightness altitude values, using the same categories.  Using the appropriate parameters of the Arizona bolide, this object lies on the border of Type I/Type II regimes. Using bulk density value of 2000 ${\rm kg/m^3}$ for Type II objects \citep{Ceplecha1988}, we obtained a diameter of $2.42 \pm 0.09$ m. 

For Type I objects, we take the density of about $3000$ kg/m$^3$ to be consistent with a weaker material as revealed by the fragmentation model. This gives a size of $2.10\pm0.08$ m.

Looking at the SkySentinel MSR observations, the data indicates strong emission in the blue filter and weak emission in the green. This suggests that the object is Fe-poor, as a significant number of Fe emission lines are in the green band. The high starting altitude of $\sim100$ km points to a low material strength object of asteroidal origin as stated above. For Fe-poor asteroids with low material strength, we adopted the bulk density value of $2500 {\rm kg/m^3}$ was from \citet{britt2003}.  Assuming spherical impactors, we calculated an initial diameter of $2.25 \pm 0.08$ m for the bolide. 

Analysis of the recovered fragments categorize the object as being an LL7 chondrite with a shock stage S0 \citep{METBulletin}.  \citet{britt2003} performed a bulk density analysis for this material type and we adopted their value of $3220 \pm 220 {\rm kg/m^3}$ which results in an pre-impact diameter for the Arizona bolide of $2.03 \pm 0.12$ m. The values of $K=0.35$ cm$^2$/g$^{2/3}$ and $\sigma=0.045$ s$^2$/km$^2$, corresponding to a density of about $2500$-$3000$ kg/m$^3$ for $\Gamma A'\sim0.7$ chosen in our fragmentation model are consistent with a weak object of this type.

The summary of results from all analyses described above is shown in Table~\ref{tab:sizes}. 

\section{Discussion and Conclusions}
We have analyzed multi-station observations of a magnitude $-20.4 \pm 0.2$ superbolide that entered Earth's atmosphere over Arizona on June 2, 2016.  The calculated deposited energy of the fireball from the light curve analysis is $0.54 \pm 0.06$ kt. This is about $12\%$ larger than the value of 0.48 kt reported by CNEOS based on data from US government sensors. Their reported peak location of $33.8\degree$N and $110.9\degree$W longitude (marked in Figure \ref{fig:trajectory}) is about 26 km off from our calculated position of the peak brightness.  The site of recovered fragments are along the line of our calculated trajectory, thus we believe that for this particular event the CNEOS reported peak brightness location is inaccurate. \\
From the deposited energy and the entry velocity of the bolide we estimate the mass of the object to be $14.8 \pm 1.7$ metric tonnes.  We calculate the initial size of the object by assuming spherical shape.  We use 3 different methods to obtain the bulk density of the meteor based on the observations of the peak altitude, end hight, and spectral emission during the ablation. \citet{Borovicka2017} noted that the method of end height analysis \citep{Ceplecha1976} for superbolides may lead to misleading results, while the use of the height of peak brightness \citep{Brown2016} to determine material strength does not provides an adequate fit.  Therefore, the true size of the object is most likely smaller than what we obtained from these methods. \\
Analysis of the MSR data points to a Fe-poor object. 
Assuming asteroidal origin, an Fe-poor composition, such as CI/CM chondrite, gives a slightly smaller diameter than the peak/end height analysis. The reason we test different methods is that in the absence of knowing the exact composition we would have to rely on these types of analyses. In this particular case, there were recovered fragments from this bolide and using the actual LL7 chondrite composition we estimate the diameter of the object to be $2.03 \pm 0.12$ m. Our spectral analysis is in agreement with this value while the peak brightness method overestimate this diameter by about 17\%. In summary, the object is most likely to be $\sim2$m in diameter with a bulk density between $2000-3000$ kg/m$^3$. \\
The calculated orbit points to an origin interior to the main belt that is nearly co-orbital to the Earth. The object did not form part of any known families, and a search for the parent body yielded no results. \citet{Dunn2013} state that a majority of Apollo-class asteroids are LL-chondrites; this object is therefore not unusual.


\section*{Acknowledgements}
The authors would like to thank J. Borovi{\v c}ka for his thorough review and helpful comments that greatly improved the clarity of the manuscript. 
We also wish to acknowledge those responsible for the operation and maintenance of the individual Nodes used in the analysis of the Spalding Allsky Camera Network data. \\
Nodes 1 and 2 - NMSU, Las Cruces, NM: Dr. Robert Wagner\\
Node 6 - Flagstaff, AZ: Steven Schoner\\
Node 14 - Los Alamos, NM: Dr. Matt Heavner\\
Node 37 - Parker, AZ: Jim Woodell (no longer operational)\\
Node 79 - Turkey Springs Observatory, Payson, AZ: Bruce Rasch\\
Node 7 - Lamy, NM: Dr. Thomas Ashcraft\\
Nodes 5 and 8 - Albuquerque, NM: Dwayne Free\\
We would also like to thank Dr. Jeremy Riousset for his helpful suggestions. 



\bibliographystyle{mnras}
\bibliography{palotai.2017.bolide}

\begin{thebibliography}{}
\makeatletter
\relax
\def\mn@urlcharsother{\let\do\@makeother \do\$\do\&\do\#\do\^\do\_\do\%\do\~}
\def\mn@doi{\begingroup\mn@urlcharsother \@ifnextchar [ {\mn@doi@}
  {\mn@doi@[]}}
\def\mn@doi@[#1]#2{\def\@tempa{#1}\ifx\@tempa\@empty \href
  {http://dx.doi.org/#2} {doi:#2}\else \href {http://dx.doi.org/#2} {#1}\fi
  \endgroup}
\def\mn@eprint#1#2{\mn@eprint@#1:#2::\@nil}
\def\mn@eprint@arXiv#1{\href {http://arxiv.org/abs/#1} {{\tt arXiv:#1}}}
\def\mn@eprint@dblp#1{\href {http://dblp.uni-trier.de/rec/bibtex/#1.xml}
  {dblp:#1}}
\def\mn@eprint@#1:#2:#3:#4\@nil{\def\@tempa {#1}\def\@tempb {#2}\def\@tempc
  {#3}\ifx \@tempc \@empty \let \@tempc \@tempb \let \@tempb \@tempa \fi \ifx
  \@tempb \@empty \def\@tempb {arXiv}\fi \@ifundefined
  {mn@eprint@\@tempb}{\@tempb:\@tempc}{\expandafter \expandafter \csname
  mn@eprint@\@tempb\endcsname \expandafter{\@tempc}}}

\bibitem[\protect\citeauthoryear{{Astropy Collaboration} et~al.,}{{Astropy
  Collaboration} et~al.}{2018}]{AstropyPaper}
{Astropy Collaboration} et~al., 2018, \mn@doi [\aj] {10.3847/1538-3881/aabc4f},
  \href {http://adsabs.harvard.edu/abs/2018AJ....156..123A} {156, 123}

\bibitem[\protect\citeauthoryear{{Bannister}, {Boucheron}  \&
  {Voelz}}{{Bannister} et~al.}{2013}]{Bannister2013}
{Bannister} S.~M.,  {Boucheron} L.~E.,   {Voelz} D.~G.,  2013, \mn@doi [\pasp]
  {10.1086/673167}, \href {http://adsabs.harvard.edu/abs/2013PASP..125.1108B}
  {125, 1108}

\bibitem[\protect\citeauthoryear{{Borovicka}}{{Borovicka}}{1990}]{borovicka1990}
{Borovicka} J.,  1990, Bulletin of the Astronomical Institutes of
  Czechoslovakia, \href {http://adsabs.harvard.edu/abs/1990BAICz..41..391B}
  {41, 391}

\bibitem[\protect\citeauthoryear{{Borovicka}, {Spurny}  \&
  {Keclikova}}{{Borovicka} et~al.}{1995}]{borovicka1995}
{Borovicka} J.,  {Spurny} P.,   {Keclikova} J.,  1995, \aaps, \href
  {http://adsabs.harvard.edu/abs/1995A%26AS..112..173B} {112, 173}

\bibitem[\protect\citeauthoryear{{Borovi{\v c}ka}}{{Borovi{\v
  c}ka}}{1992}]{Borovicka1992}
{Borovi{\v c}ka} J.,  1992, Publications of the Astronomical Institute of the
  Czechoslovak Academy of Sciences, \href
  {http://adsabs.harvard.edu/abs/1992PAICz..79.....B} {79}

\bibitem[\protect\citeauthoryear{{Borovi{\v c}ka} et~al.,}{{Borovi{\v c}ka}
  et~al.}{2013}]{Borovicka2013}
{Borovi{\v c}ka} J.,  et~al., 2013, \mn@doi [Meteoritics and Planetary Science]
  {10.1111/maps.12078}, \href
  {http://adsabs.harvard.edu/abs/2013M%26PS...48.1757B} {48, 1757}

\bibitem[\protect\citeauthoryear{{Borovi{\v c}ka}, {Spurn{\'y}}, {Grigore}  \&
  {Svore{\v n}}}{{Borovi{\v c}ka} et~al.}{2017}]{Borovicka2017}
{Borovi{\v c}ka} J.,  {Spurn{\'y}} P.,  {Grigore} V.~I.,   {Svore{\v n}} J.,
  2017, \mn@doi [Planetary and Space Science]
  {https://doi.org/10.1016/j.pss.2017.02.006}, 143, 147

\bibitem[\protect\citeauthoryear{{Britt} \& {Consolmagno}}{{Britt} \&
  {Consolmagno}}{2003}]{britt2003}
{Britt} D.~T.,  {Consolmagno} G.~J.,  2003, \mn@doi [Meteoritics and Planetary
  Science] {10.1111/j.1945-5100.2003.tb00305.x}, \href
  {http://adsabs.harvard.edu/abs/2003M%26PS...38.1161B} {38, 1161}

\bibitem[\protect\citeauthoryear{{Brown}, {Spalding}, {ReVelle}, {Tagliaferri}
  \& {Worden}}{{Brown} et~al.}{2002}]{brown2002nature}
{Brown} P.,  {Spalding} R.~E.,  {ReVelle} D.~O.,  {Tagliaferri} E.,   {Worden}
  S.~P.,  2002, \mn@doi [\nat] {10.1038/nature01238}, \href
  {http://adsabs.harvard.edu/abs/2002Natur.420..294B} {420, 294}

\bibitem[\protect\citeauthoryear{{Brown}, {Wiegert}, {Clark}  \&
  {Tagliaferri}}{{Brown} et~al.}{2016}]{Brown2016}
{Brown} P.,  {Wiegert} P.,  {Clark} D.,   {Tagliaferri} E.,  2016, \mn@doi
  [\icarus] {10.1016/j.icarus.2015.11.022}, \href
  {http://adsabs.harvard.edu/abs/2016Icar..266...96B} {266, 96}

\bibitem[\protect\citeauthoryear{{Ceplecha}}{{Ceplecha}}{1987}]{Ceplecha1987}
{Ceplecha} Z.,  1987, Bulletin of the Astronomical Institutes of
  Czechoslovakia, \href {http://adsabs.harvard.edu/abs/1987BAICz..38..222C}
  {38, 222}

\bibitem[\protect\citeauthoryear{{Ceplecha}}{{Ceplecha}}{1988}]{Ceplecha1988}
{Ceplecha} Z.,  1988, Bulletin of the Astronomical Institutes of
  Czechoslovakia, \href {http://adsabs.harvard.edu/abs/1988BAICz..39..221C}
  {39, 221}

\bibitem[\protect\citeauthoryear{{Ceplecha} \& {McCrosky}}{{Ceplecha} \&
  {McCrosky}}{1976}]{Ceplecha1976}
{Ceplecha} Z.,  {McCrosky} R.~E.,  1976, \mn@doi [\jgr]
  {10.1029/JB081i035p06257}, \href
  {http://adsabs.harvard.edu/abs/1976JGR....81.6257C} {81, 6257}

\bibitem[\protect\citeauthoryear{{Ceplecha} \& {Revelle}}{{Ceplecha} \&
  {Revelle}}{2005}]{Ceplecha2005}
{Ceplecha} Z.,  {Revelle} D.~O.,  2005, \mn@doi [Meteoritics and Planetary
  Science] {10.1111/j.1945-5100.2005.tb00363.x}, \href
  {http://adsabs.harvard.edu/abs/2005M%26PS...40...35C} {40, 35}

\bibitem[\protect\citeauthoryear{{Ceplecha}, {Borovi{\v c}ka}, {Elford},
  {Revelle}, {Hawkes}, {Porub{\v c}an}  \& {{\v S}imek}}{{Ceplecha}
  et~al.}{1998}]{Ceplecha1998}
{Ceplecha} Z.,  {Borovi{\v c}ka} J.,  {Elford} W.~G.,  {Revelle} D.~O.,
  {Hawkes} R.~L.,  {Porub{\v c}an} V.,   {{\v S}imek} M.,  1998, \mn@doi [\ssr]
  {10.1023/A:1005069928850}, \href
  {http://adsabs.harvard.edu/abs/1998SSRv...84..327C} {84, 327}

\bibitem[\protect\citeauthoryear{{Drummond}}{{Drummond}}{1981}]{Drummond1981}
{Drummond} J.~D.,  1981, \mn@doi [\icarus] {10.1016/0019-1035(81)90020-8},
  \href {http://adsabs.harvard.edu/abs/1981Icar...45..545D} {45, 545}

\bibitem[\protect\citeauthoryear{{Dunn}, {Burbine}, {Bottke}  \&
  {Clark}}{{Dunn} et~al.}{2013}]{Dunn2013}
{Dunn} T.~L.,  {Burbine} T.~H.,  {Bottke} W.~F.,   {Clark} J.~P.,  2013,
  \mn@doi [\icarus] {10.1016/j.icarus.2012.11.007}, \href
  {http://adsabs.harvard.edu/abs/2013Icar..222..273D} {222, 273}

\bibitem[\protect\citeauthoryear{{Garvie}}{{Garvie}}{2017}]{METBulletin}
{Garvie} L.,  2017, Meteoritical Bulletin: Entry for Dishchii'bikoh, \url
  {https://www.lpi.usra.edu/meteor/metbull.php?code=65525}

\bibitem[\protect\citeauthoryear{{Jopek}}{{Jopek}}{1993}]{Jopek1993}
{Jopek} T.~J.,  1993, \mn@doi [\icarus] {10.1006/icar.1993.1195}, \href
  {http://adsabs.harvard.edu/abs/1993Icar..106..603J} {106, 603}

\bibitem[\protect\citeauthoryear{{Mirametrics Inc.}}{{Mirametrics
  Inc.}}{2017}]{Mira64}
{Mirametrics Inc.} 2017, Mira Pro x64, \url {https://www.mirametrics.com}

\bibitem[\protect\citeauthoryear{{Rein} \& {Liu}}{{Rein} \&
  {Liu}}{2012}]{rein2011}
{Rein} H.,  {Liu} S.~F.,  2012, \mn@doi [\aap] {10.1051/0004-6361/201118085},
  \href {https://ui.adsabs.harvard.edu/\#abs/2012A&A...537A.128R} {537, A128}

\makeatother
\end{thebibliography}





\bsp	
\label{lastpage}
\end{document}